\documentclass[lettersize,journal]{IEEEtran}
\usepackage{amsmath,amsfonts}
\usepackage{algorithmic}
\usepackage{algorithm}
\usepackage{array}
\usepackage[caption=false,font=normalsize,labelfont=sf,textfont=sf]{subfig}
\usepackage{textcomp}
\usepackage{stfloats}
\usepackage{url}
\usepackage{verbatim}
\usepackage{graphicx}
\usepackage{cite}
\usepackage{booktabs}
\usepackage{multirow}
\usepackage{makecell} 
\begin{document}

\title{Physics-driven AI for Channel Estimation in Cellular Network}

\author{Xiaoqian Qi,~\IEEEmembership{Member,~IEEE,} Haoye Chai,~\IEEEmembership{Member,~IEEE,} Yong Li,~\IEEEmembership{Member,~IEEE}}

\markboth{Journal of \LaTeX\ Class Files,~Vol.~14, No.~8, August~2021}%
{Shell \MakeLowercase{\textit{et al.}}: A Sample Article Using IEEEtran.cls for IEEE Journals}


\maketitle

\begin{abstract}
In cellular mobile networks, wireless channel quality (CQ) is a crucial factor in determining communication performance and user's network experience. Accurately predicting CQ based on real environmental characteristics, specific base station configurations and user trajectories can help network operators optimize base station deployment, improving coverage and capacity. The Received Signal Reference Power (RSRP) and Signal-to-Interference-plus-Noise Ratio (SINR) of user equipment (UE) are key indicators of CQ in wireless communication. However, existing researches have limitations in terms of generation accuracy. Regression methods such as statistical inference and random forests fail to effectively capture the unique characteristics of wireless environments; theoretical derivations relying on specific communication protocols lack generalization capability; data-driven machine learning (ML) methods like Long Short-Term Memory (LSTM) Network often suffer from a lack of interpretability. To overcome these limitations, we propose physics-informed diffusion models, which accurately generate RSRP and SINR at UE based on the wireless environment, base station configurations, and user trajectories. The model adopts a modular and end-to-end design, employing a teacher-student framework to achieve knowledge distillation. This method integrates expert knowledge into the training of diffusion models, enhancing both the interpretability and accuracy, while also facilitating faster convergence of the model parameters. Furthermore, it allows for self-adaptation in various scenarios through few-shot learning. This approach provides valuable guidance for optimizing base station deployment, predicting user network experience, and building real-world simulators.
\end{abstract}

\begin{IEEEkeywords}
Channel Quality Generation, Diffusion Models, Knowledge Distillation.
\end{IEEEkeywords}

\section{Introduction}

\IEEEPARstart{C}{hannel} quality (CQ) is a key factor limiting the performance of wireless communication systems. CQ determines both the efficiency and reliability of communication, and directly impacts the user's network experience. In wireless environments, unpredictable interference and random noise introduce fading and delay, leading to data loss and errors after reception and demodulation. Therefore, accurately predicting wireless channel quality in specific scenarios is a critical step in optimizing wireless communication. For cellular networks, operators must strategically plan base station layout and configuration before deployment to optimize coverage and signal quality, ensuring stable communication services across different areas. The Received Signal Reference Power (RSRP) and Signal-to-Interference-plus-Noise Ratio (SINR) are key metrics for evaluating network coverage and communication quality. By estimating RSRP and SINR under specific wireless environments, operators can more accurately plan base station deployments and select appropriate Modulation and Coding Schemes (MCS), thereby optimizing resource utilization and improving network performance. Real-time channel estimation relies on decoding periodically transmitted pilots to obtain and feedback Channel State Information (CSI), which is not applicable before network deployment. Therefore, it is essential to develop a predictive paradigm that does not rely on real-time feedback.

We define the problem as accurately generating RSRP and SINR, which represent channel quality, based on specific scenarios. Specifically, in real-world wireless environments that include streets, buildings, and terrain features, given a cellular network configuration that includes base station locations and parameters, as well as multiple users' continuous trajectories, we generate the RSRP and SINR for each user at every time point. This is essentially a conditional generation problem, where the conditional information formed by the environment, base stations, and trajectories determines the values of RSRP and SINR. In static, stationary channels, RSRP and SINR can often be calculated precisely through theoretical models. However, when randomness and non-stationarity are introduced by real environments, RSRP and SINR deviate from their theoretical values under static and stationary conditions, producing an environment-dependent gain. Based on this mechanism, a reasonable approach is to construct a conditional generation model that takes base stations and trajectories as fundamental conditions and learns the environmental features through the model, thereby generating accurate RSRP and SINR values.

Existing approaches for predicting RSRP and SINR can be broadly classified into three categories: physics-based propagation models, regression analysis with feature engineering, and data-driven methods based on deep learning. ......

However, these approaches have limitations in terms of prediction accuracy and generalizability. Specifically, physics-based propagation models can predict channels under certain scenarios based on specific communication protocols, but these models heavily rely on prior knowledge of a single protocol, making it difficult to generalize to complex environments. Regression analysis with feature engineering constrains the relationships between physical quantities to specific regression models and fails to capture unobserved environmental characteristics. Meanwhile, data-driven deep learning (DL) methods, such as LSTM, lack interpretability, and simple models are insufficient in extracting underlying mechanisms from environmental features and physical models.

To overcome the limitations of existing research, we propose physics-informed diffusion models based on the idea of digital twin of wireless network environment. This framework uses conditional diffusion models as the core architecture for generation and embeds wireless transmission models from various scenarios as physical knowledge into the diffusion model training process via knowledge distillation. \textbf{First}, for the conditional diffusion model, we consider the fundamental factors affecting RSRP and SINR, including the user's position at each time point, and the connected base station's location, height, frequency, and transmission power, as the primary conditions. \textbf{Second}, for the wireless transmission model, we categorize the environment into three types: urban, suburban, and rural, and use measurement-based transmission models specific to each scenario to calculate the initial theoretical values of RSRP, thereby forming concrete representations of physical knowledge. \textbf{Third}, we embed physical knowledge into the training process of the diffusion model using a knowledge distillation approach. We adopt a Teacher-Student training framework, where the diffusion model is first pre-trained using the initial RSRP values based on physical models, followed by fine-tuning on real-world data to further capture the gains introduced by environmental features. \textbf{In addition}, the model employs an end-to-end design, utilizing the diffusion model to simultaneously generate both RSRP and SINR. Since SINR is influenced by more random factors, such as interference and noise, but maintains a close correlation with RSRP, we treat RSRP and SINR as two attributes of the target vector and design a Two-dimensional (2D) Transformer to extract features simultaneously along both the temporal and attribute dimensions. This design ensures that the generated SINR not only relies on the more accurately modeled RSRP, but also captures the dynamic nature of digital twins in time-series data, while avoiding the error accumulation typically associated with modular designs. This ultimately improves the model's generation accuracy and real-world applicability.

\begin{itemize}
    \item We construct a digital twin of wireless network environments with a conditional diffusion model at its core, enabling controlled generation of each user's RSRP and SINR at every time point based on the environment, base station configurations, and user trajectories. This allows for accurate prediction of channel quality in known wireless environments. To the best of our knowledge, this is the first time that a diffusion model has been applied to the wireless channel prediction problem.

    \item We introduce physical knowledge into the training process of the diffusion model using a knowledge distillation approach, enhancing the interpretability of the DL network when simulating wireless transmission processes. This design achieves complementary strengths between physical knowledge and DL networks. The incorporation of physical knowledge helps the diffusion models identify the relationships between target and conditional variables, while the feature extraction capabilities of the diffusion models and the 2D Transformer compensate for the shortcomings of physical knowledge in capturing environmental characteristics.

    \item Our model enables the end-to-end simultaneous generation of RSRP and SINR, thereby avoiding error propagation and accumulation between modules. With the help of the 2D Transformer, this design also captures the temporal features of both RSRP and SINR, as well as the correlation between the two, further improving the generation accuracy and the model's expressiveness.
    
\end{itemize}
\section{System Model}
\subsection{Wireless Scenario}

\begin{figure}[tb]
\centering
\includegraphics[width=\linewidth]{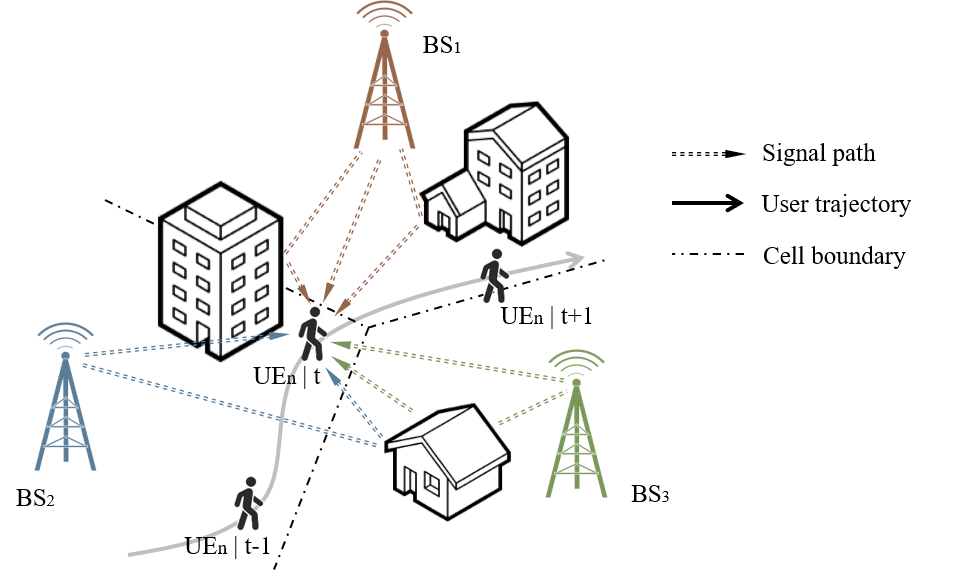}
\caption{System model: cellular mobile communication.}
\vspace{-4mm}
\label{Fig_syst}
\end{figure}

We consider a 5G cellular communication scenario that utilizes a multi-cell Multiple-Input Multiple-Output (MIMO) system, where each Next-Generation-Node-B (gNodeB) base station (BS) is equipped with multiple antennas to provide high-speed data services to multiple user equipment (UE). It is assumed that the frequency range is selected from the low band (Sub-1GHz) to the mid-to-high band (Sub-6GHz). For example, China Mobile commonly uses three fixed frequency points: 700MHz, 2.6GHz, and 4.9GHz. To address the challenges posed by the high density of buildings and users in urban environments, the system typically adopts ultra-dense networking (UDN), where each base station covers a relatively small area, and multiple base stations coordinate via frequency reuse. We assume that each UE receives signals from neighboring base stations and selects the one with the highest received signal power (PR) as target for connection. 

As shown in Fig. \ref{Fig_syst}, the area of interest (AOI) consists of multiple adjacent cells, with a total of $N \geq 1$ BSs and $M \geq 1$ users, as well as other entities such as buildings and trees. Users may experience issues such as signal blocking due to building obstructions and multipath effects caused by signal reflections. Without loss of generality, we assume that user equipment $\text{UE}_m$ can receive signals from base stations $\{BS_i | i=1, 2, ..., s_m\}$ at time $t$, where $m \in \mathcal{M}=\{1,2,...,M\}$, $t \in \mathcal{T}=\{1,2,...,T\}$. $K_{i,m}^{(t)} > 0$ is the number of reflected paths, $\{\Theta_1, \Theta_2, ..., \Theta_{K_{i,m}^{(t)}}\}$, between base station ${\text{BS}_i}$ and $\text{UE}_m$. We assume that the communication protocol is reliable, where real-time CSI estimation feedback can effectively account for the impact of multipath effects on the received signal. Thus, the multipath signals, $\pmb{X}_{i,m}^{(t)}=[\pmb{x}_{i,m,1}^{(t)}, \pmb{x}_{i,m,2}^{(t)}, ..., \pmb{x}_{i,m,K_{i,m}}^{(t)}]$ received by $\text{UE}_m$ at time $t$ from the same base station $\{\text{BS}_i\}$ can all be effectively utilized. The corresponding PR is expressed as
\begin{equation}
    \text{PR}_{i,m}^{(t)}=\sum_{k=1}^{K_{i,m}^{(t)}}|\pmb{x}_{i,m,k}^{(t)}|^2=||\pmb{X}_{i,m}^{(t)}||^2_2.
\label{eq_PR}
\end{equation}

$\text{UE}_m$ will choose the BS with the highest receive power as the target BS to connect, that
\begin{equation}
\begin{aligned}
    \text{RSRP}_{m}^{(t)} &= \underset{i \in \{1, 2, ..., s_m\}}{\max} \text{PR}_{i,m}^{(t)},\\
    s_{\text{ta}} &= \underset{i \in \{1, 2, ..., s_m\}}{\arg\max} \text{PR}_{i,m}^{(t)},
\end{aligned}
\label{eq_RSRP}
\end{equation}
where $s_{\text{ta}} \in \{1, 2, ..., s_m\}$ is the index of the target BS.

According to the above analysis, since the user can receive transmission signals from multiple base stations in neighboring cells, and these signals operate at frequencies belonging to the set of frequencies $\mathcal{F}$ determined by the operator's communication protocol, co-channel interference will occur between the non-serving base stations and the serving base station operating at the same frequency. Assuming the frequencies of the $N$ base stations are ${f_1, f_2, ..., f_N} \in \mathcal{F}^N$, the co-channel interference power from neighboring base stations to the target base station's signal can be calculated as
\begin{equation}
    \pmb{\mathcal{I}}_{m}^{(t)} = \sum_{\substack{i=1 \\ i \neq s_{\text{ta}}}}^{s_m} \text{PR}_{i,m}^{(t)} \cdot \mathbf{1}\{f_i = f_{s_{\text{ta}}}\}.
\label{eq_Interference}
\end{equation}

The SINR at the receiver is defined as the ratio of the RSRP to the sum power of interference and noise, that
\begin{equation}
    \text{SINR}_{m}^{(t)} = \frac{\text{RSRP}_{m}^{(t)}}{\pmb{\mathcal{I}}_{m}^{(t)}+\pmb{\mathcal{N}}_{m}^{(t)}},
\label{eq_SINR}
\end{equation}
where $\pmb{\mathcal{N}}_{m}^{(t)}$ is the power of random noise.

\subsection{Wireless Propagation Model}
\label{subsec_WPM}
Typically, the transmission power (PT) of BS is manually set. For example, in the commonly used 3GPP protocol for 5G communication, the maximum transmission power of a macro base station (Macro-BS) usually ranges from 40W to several hundred watts, while the transmission power of a micro base station (Micro-BS) typically ranges from 1W to 5W to accommodate smaller coverage areas. However, the received power of mobile user equipment is generally no more than 0.01 mW, and in cases of weak signals or longer distances, the received power may drop to as low as 1 pW or even lower. A significant amount of energy is lost during signal transmission due to absorption by the medium, as well as reflection and scattering off the surfaces of objects. Therefore, to accurately estimate the received power, we need to model the propagation of signals in the wireless environment.

Assuming the base station transmit power is $P_T$, the transmit antenna gain is $G_t$, the receive antenna gain is $G_r$, and the path loss is $PL$, the received power can typically be expressed as
\begin{equation}
    \text{PR}=\text{PT}+G_r+G_t-\text{PL},
\label{eq_PR}
\end{equation}
where PT, $G_t$, and $G_r$ are all controllable device parameters, so the key to estimation lies in accurately calculating the path loss in different environments.

The \textbf{Free Space Propagation Model (FSPM)} is one of the most fundamental propagation models in wireless communication. It is used to describe signal attenuation as radio waves propagate through free space. The model assumes that there are no obstacles, reflections, diffractions, or scattering effects along the propagation path, making it primarily suitable for propagation analysis in ideal environments. In FSPM, the path loss is given by
\begin{equation}
\begin{split}
    &\text{PL}(\text{W}) = \left(\frac{4 \pi d f_c}{c}\right)^2, \text{ or}\\
    &\text{PL}(\text{dB}) = 20 \log_{10}(d) + 20 \log_{10}(f_c) + 20 \log_{10}\left(\frac{4\pi}{c}\right),
\end{split}
\end{equation}
where $d$ is the distance between the transmitter and the receiver (in m), $f_c$ is the frequency of the signal (in Hz), and $c=3\times 10^8$ m/s is the speed of light.

However, most mobile communication systems operate in complex propagation environments, where accurate path loss modeling cannot be fully captured by FSPL. To address this, researchers have developed numerous \textbf{Measurement-based Propagation Models}, designed to accurately simulate path loss across a variety of typical wireless environments. Table \ref{Tab_Hata} presents the two models used in this paper, the Hata model~\cite{Hata} (applicable in $150\sim1500$ MHz range) and the WINNER II model~\cite{WINNERII} (applicable in the $2\sim6$ GHz range), along with their path loss formulas for different environments.

\begin{table*}[!h]
\centering
\small
\caption{Path Loss in Urban, Suburban and Rural Scenarios According to Hata Model and WINNER II Model.}
\begin{tabular}{clc}
\toprule
Model & Path Loss (dB) & Range of $f_c$ \\
\midrule
\multirow{9}{*}{Hata} 
     & $\text{PL}_{\text{urban}}=69.55+26.16\log_{10}(f_c)-13.82\log_{10}(h_t)-\alpha(h_r)+[44.9-6.55\log_{10}(h_t)]\log_{10}(d)$ & \multirow{9}{*}{$150\sim1500$ MHz} \\
     & for big city, $\alpha(h_r)=3.2[\log_{10}(11.75h_r)]^2-4.97$ & \\
     & for middle and small city, $\alpha(h_r)=[1.1\log_{10}(f_c)-0.7]h_r-[1.56\log_{10}(f_c)-0.8]$ & \\
\cmidrule(lr){2-2}     
     & $\text{PL}_{\text{suburb}}=\text{PL}_{\text{urban}}-2[\log_{10}(f_c/20)]^2-5.4$ & \\
\cmidrule(lr){2-2}
     & $\text{PL}_{\text{rural}}=\text{PL}_{\text{urban}}-4.78[\log_{10}(f_c)]^2+18.33\log_{10}(f_c)-K$ & \\
     & $K$ is an environmental factor, ranging from 35.94 (rural) to 40.94 (desert). & \\
     & & \\
     & \textit{Hint: $f_c$ in} MHz, \textit{$h_t$ in} m, \textit{$h_r$ in} m, and \textit{$d$ in} km. & \\
\midrule
\multirow{6}{*}{WINNER II} 
     & $\text{PL}_{\text{urban}} = 40 \log_{10}(d_l) + 9.45 - 17.3 \log_{10}(h_t) - 17.3 \log_{10}(h_r) + 2.7 \log_{10}(f_c/5)$ & \multirow{6}{*}{$2\sim6$ GHz} \\
\cmidrule(lr){2-2}     
     & $\text{PL}_{\text{suburb}} = 40 \log_{10}(d_l) + 11.65 - 16.2 \log_{10}(h_t) - 16.2 \log_{10}(h_r) + 3.8 \log_{10}\left(f_c/5\right)$ & \\
\cmidrule(lr){2-2}
     & $\text{PL}_{\text{rural}} = 40 \log_{10}(d_l) + 10.5 - 18.5 \log_{10}(h_t) - 18.5 \log_{10}(h_r) + 1.5 \log_{10}\left(f_c/5\right)$ & \\
     & & \\
     & \textit{Hint: $f_c$ in} GHz, \textit{$h_t$ in} m, \textit{$h_r$ in} m, and \textit{$d$ in} m. & \\
\bottomrule
\end{tabular}
\label{Tab_Hata}
\end{table*}

\subsection{Problem Formulation}
In this paper, we implement the generation of RSRP and SINR under specific wireless scenarios using a physics-informed diffusion model. The wireless propagation model in Section \ref{subsec_WPM} provides the physical modeling of the wireless channel. The generation of RSRP and SINR is essentially a conditional time-series generation problem. For user equipment $\text{UE}_m$ in AOI, considering a time range of length $T$, we aim to estimate
\begin{align}
    \pmb{\text{RSRP}}_m^{1\times T}&=[\text{RSRP}_{m}^{(1)}, \text{RSRP}_{m}^{(2)},...,\text{RSRP}_{m}^{(T)}], \\
    \pmb{\text{SINR}}_m^{1\times T}&=[\text{SINR}_{m}^{(1)}, \text{SINR}_{m}^{(2)},...,\text{SINR}_{m}^{(T)}].
\end{align}
Define $\Gamma_m \in \{\text{urban}, \text{suburb},\text{rural}\}$ as the type of AOI, the conditions for the generation model are defined as the following quintuple
\begin{equation}
    \pmb{Con}_m^{1\times T}=(\pmb{d}_m^{1\times T},\pmb{h_r}_m^{1\times T},\pmb{f}_m^{1\times T},\pmb{P_r}_m^{1\times T},\Gamma_m),
\end{equation}
where
\begin{equation}
    \pmb{\alpha}_m^{1\times T}=[\alpha_{m}^{(1)}, \alpha_{m}^{(2)}, ...,\alpha_{m}^{(T)}], \quad \alpha \in \{d,h_r,f_,P_r\}.
\end{equation}
Therefore, let the physics-informed generation model be denoted as $\pmb{\Phi}^{Pi}$, and we define the generation problem as fitting the joint conditional distribution of the target estimators:
\begin{equation}
    p(\pmb{\text{RSRP}}_m^{1\times T},\pmb{\text{SINR}}_m^{1\times T}|\pmb{Con}_m^{1\times T})=\pmb{\Phi}^{Pi}(\pmb{Con}_m^{1\times T}).
\end{equation}
\section{Physics-informend Diffusion Models for RSRP and SINR Gereration}
\subsection{Conditional Diffusion Models}
\label{CSDI}
Diffusion models, originally proposed by Sohl-Dickstein \textit{et al.}~\cite{DBLP}, are novel deep generative models having outstanding performance in tasks like image generation. It works by  adding noise to the data step by step during the \textbf{forward process} and recovering data by estimating the noise in the \textbf{reverse process}. This method uses variational inference to ensure that samples resemble the actual data distribution over a finite number of steps. In order to achieve controlled data generation under specific conditions using the diffusion model. Tashiro \textit{et al.}~\cite{CSDI} extended this with a conditional diffusion model, aiming to generate data conditioned on specific variables. Given the target space $\pmb{\mathcal{X}}^{ta}$ and the condition space $\pmb{\mathcal{X}}^{co}$, the goal is to approximate the distribution $q(\pmb{x}^{ta}_0)$ by learning a model for the conditional distribution $p_\theta(\pmb{x}^{ta}_0|\pmb{x}_0^{co})$. Here, $\pmb{x}^{ta}_t \in \pmb{\mathcal{X}}^{ta}$ represents the latent variables, while $\pmb{x}^{co}_0 \in \pmb{\mathcal{X}}^{co}$ provides the conditioning information.
\begin{equation}
\begin{aligned}
    q(\pmb{x}^{ta}_{1:T}|\pmb{x}^{ta}_0) &:= \prod_{t=1}^T q(\pmb{x}^{ta}_t|\pmb{x}^{ta}_{t-1}), \\
    q(\pmb{x}^{ta}_t|\pmb{x}^{ta}_{t-1}) &:= \mathcal{N}(\sqrt{1-\beta_t}\pmb{x}^{ta}_{t-1}, \beta_t \pmb{I}),
\end{aligned}
\label{eq2}
\end{equation}
where $\beta_t$ reflects the intensity of added noise. Define $\hat{\alpha}_t := 1-\beta_t$, $\alpha_t := \prod_{i=1}^t \hat{\alpha}_t$, according to the recursion by Ho \textit{et al.}~\cite{DDPM}, $\pmb{x}^{ta}_t$ can be represented as $\pmb{x}^{ta}_t = \sqrt{\alpha_t}\pmb{x}^{ta}_0 + (1-\alpha_t)\epsilon$, where $\epsilon \sim \mathcal{N}(\pmb{0}, \pmb{I})$. 

\begin{figure*}[!h]
\centering
\includegraphics[width=0.7\linewidth]{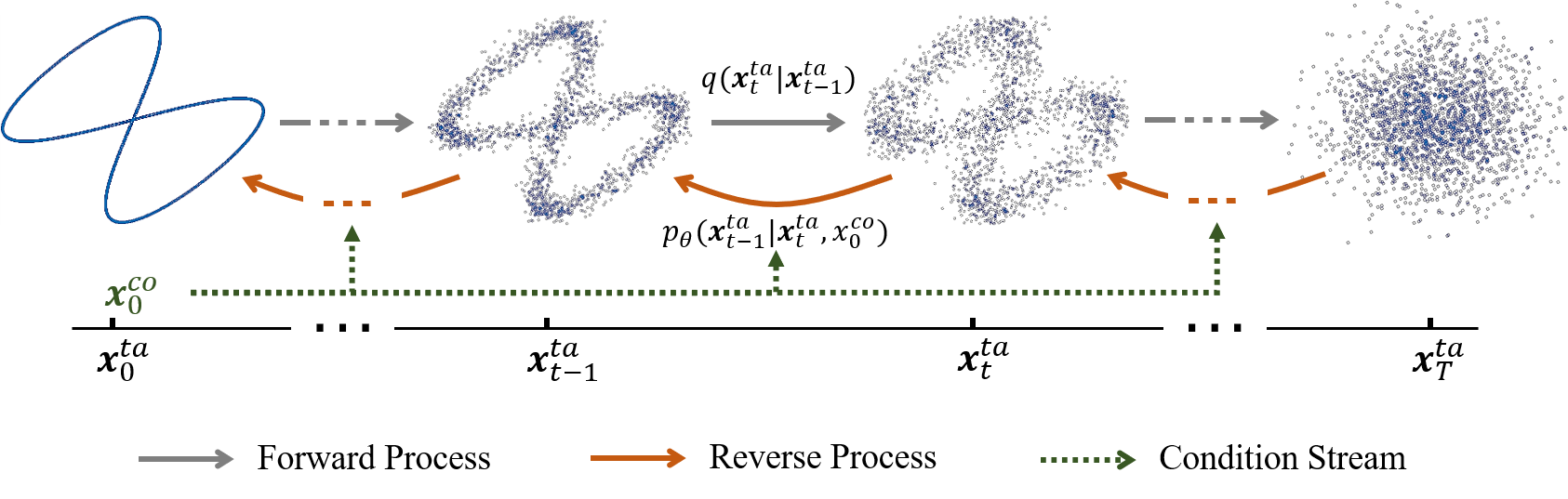}
\caption{Forward process and reverse process of conditional diffusion models.}
\vspace{-4mm}
\label{Fig2}
\end{figure*}

Conversely, the reverse process is the step-by-step restoration of $x^{ta}_0$ from the noise. It can be represented as the following Markov chain:
\begin{equation}
\begin{aligned}
    &p_\theta(\pmb{x}^{ta}_{0:T}|\pmb{x}_0^{co}) := p(\pmb{x}^{ta}_T) \prod_{t=1}^T p_\theta(\pmb{x}^{ta}_{t-1}|\pmb{x}^{ta}_t, \pmb{x}_0^{co}), \\
    &p_\theta(\pmb{x}^{ta}_{t-1}|\pmb{x}^{ta}_t, \pmb{x}_0^{co}) := \mathcal{N}(\pmb{x}^{ta}_{t-1}; \mu_\theta(\pmb{x}^{ta}_t, t|\pmb{x}_0^{co}), \sigma_\theta(\pmb{x}^{ta}_t, t|\pmb{x}_0^{co})\pmb{I}),
\end{aligned}
\label{eq3}
\end{equation}
where $\pmb{x}^{ta}_T \sim \mathcal{N}(\pmb{0}, \pmb{I})$, $\mu_\theta$ and $\sigma_\theta$ are trainable parameters, representing the mean and variance of the model recursive distribution $p_\theta(\pmb{x}^{ta}_{t-1}|\pmb{x}^{ta}_t, \pmb{x}_0^{co})$. According to \cite{CSDI},
\begin{equation}
    \mu_\theta(\pmb{x}^{ta}_t, t) = \frac{1}{\alpha_t}\left(\pmb{x}^{ta}_t - \frac{\beta_t}{\sqrt{1-\alpha_t}}\pmb{\epsilon}_\theta(\pmb{x}^{ta}_t, t|\pmb{x}^{co}_0)\right),
    \label{eq4}
\end{equation}
\begin{equation}
    \sigma_\theta(\pmb{x}^{ta}_t, t|\pmb{x}_0^{co}) = \widetilde{\beta}_t^{1/2} \quad \text{where} \quad \widetilde{\beta}_t = \begin{cases}
    \frac{1-\alpha_{t-1}}{1-\alpha_t}\beta_t & \text{if } t > 1 \\
    \beta_1 & \text{if } t = 1.
    \end{cases}
    \label{eq5}
\end{equation}
where $\pmb{\epsilon}_\theta$ is the estimation of noise, which is named as the conditional denoising function. Eq. (\ref{eq4}) and Eq. (\ref{eq5}) indicate that the estimation of the model distribution $p_\theta(\pmb{x}^{ta}_{0:T}|\pmb{x}_0^{co})$ is actually the estimation of $\pmb{\epsilon}_\theta$. We define the conditional denoising function $\pmb{\epsilon}_\theta : (\pmb{\mathcal{X}}^{ta} \times \mathbb{R}|\pmb{\mathcal{X}}^{co}) \rightarrow \mathcal{X}^{ta}$ with $\pmb{x}_0^{co}$ as the conditional input, for estimating the noise vector applied to $\pmb{x}_t$. Figure \ref{Fig2} illustrates the forward and reverse process of conditional diffusion models. The data generation is achieved from the reverse process with random noise as input.
Ho \textit{et al.} in~\cite{DDPM} formulated the following optimization problem for the estimation of $\pmb{\epsilon}_\theta$:
\begin{equation}
\begin{aligned}
    &\min_\theta \mathcal{L}(\theta) := \min_\theta \mathbb{E}_{\pmb{x}^{ta}_0 \sim q(\pmb{x}^{ta}_0), \pmb{\epsilon} \sim \mathcal{N}(\pmb{0},\pmb{I}), t} \left\| \epsilon - \pmb{\epsilon}_\theta(\pmb{x}^{ta}_t, t|\pmb{x}^{co}_0) \right\|_2^2, \\
    &\text{where} \quad \pmb{x}^{ta}_t = \sqrt{\alpha_t}\pmb{x}^{ta}_0 + (1-\alpha_t)\pmb{\epsilon}.
\end{aligned}
\label{eq6}
\end{equation}

\subsection{Physics Informing and Overall Framework}
The conditional diffusion model in Section \ref{CSDI} does not explicitly model the mathematical relationship between the condition $\pmb{x}_0^{co}$ and the target vector $\pmb{x}^{ta}$, meaning that the model's learning of the wireless propagation process is entirely data-driven. According to machine learning training principles, if the model is exposed to data with significant interference and noise, it may initially direct the physical modeling in incorrect and non-interpretable directions during the first few backpropagation steps. This would slow the convergence rate of the parameters and prevent the model from achieving optimal performance in scenarios outside the training set.

To address this issue, we propose a two-stage training paradigm inspired by the Teacher-Student framework in knowledge distillation, introducing physical knowledge into the training process of the diffusion model. The overall framework of our model and its training method is shown in Fig. \ref{Fig_syst}. Our architecture consists of three components: input data, the physical model, and the generative diffusion model. The input data (left side of Fig. \ref{Fig_syst}) represents the conditional information, which can be expressed as the quintuple $\pmb{Con}_m^{1\times T} = (\pmb{d}_m^{1\times T}, \pmb{h_r}_m^{1\times T}, \pmb{f}_m^{1\times T}, \pmb{P_r}_m^{1\times T}, \Gamma_m)$, consisting of the distance between the user and the connected base station over $T$ time steps, the height of the base station, frequency, transmission power, and the type of Area of Interest (AOI). We choose RSRP, which has a well-defined physical model, as the carrier of knowledge while simultaneously generating both RSRP and SINR. Specifically, we first use the wireless transmission model in Section \ref{subsec_WPM}, as shown in the bottom-right of Fig. \ref{Fig_syst}, to calculate the theoretical values of RSRP, $\hat{\textbf{RSRP}}_m^{1\times T}$, in the corresponding scenario. Then, we incorporate $\hat{\textbf{RSRP}}_m^{1\times T}$ as part of the training data.

As shown in the top-right of Fig. \ref{Fig_syst}, the generation model produces a two-dimensional vector, where the target vector is formed by concatenating RSRP and SINR. RSRP and SINR are treated as two attributes of the target vector. In the first phase, referred to as the \textbf{Teacher Forcing} phase (left side of the top-right in Fig. \ref{Fig_syst}), we use $\hat{\textbf{RSRP}}_m^{1\times T}$ as the training label, allowing the model to learn the theoretical relationships between RSRP and the conditional inputs. The second phase is the \textbf{Student Forcing} phase (right side of the top-right in Fig. \ref{Fig_syst}, where the pre-trained model from the Teacher Forcing phase is fine-tuned using the real $\textbf{RSRP}_m^{1\times T}$ as the label to further capture environmental features that the physical model cannot describe. Since the theoretical values of SINR are subject to significant randomness due to noise and interference, we use the real values $\textbf{SINR}_m^{1\times T}$ as the label in both phases. This approach enables SINR to learn the complex environmental characteristics over more training iterations.

\begin{figure*}[!h]
\centering
\includegraphics[width=0.9\linewidth]{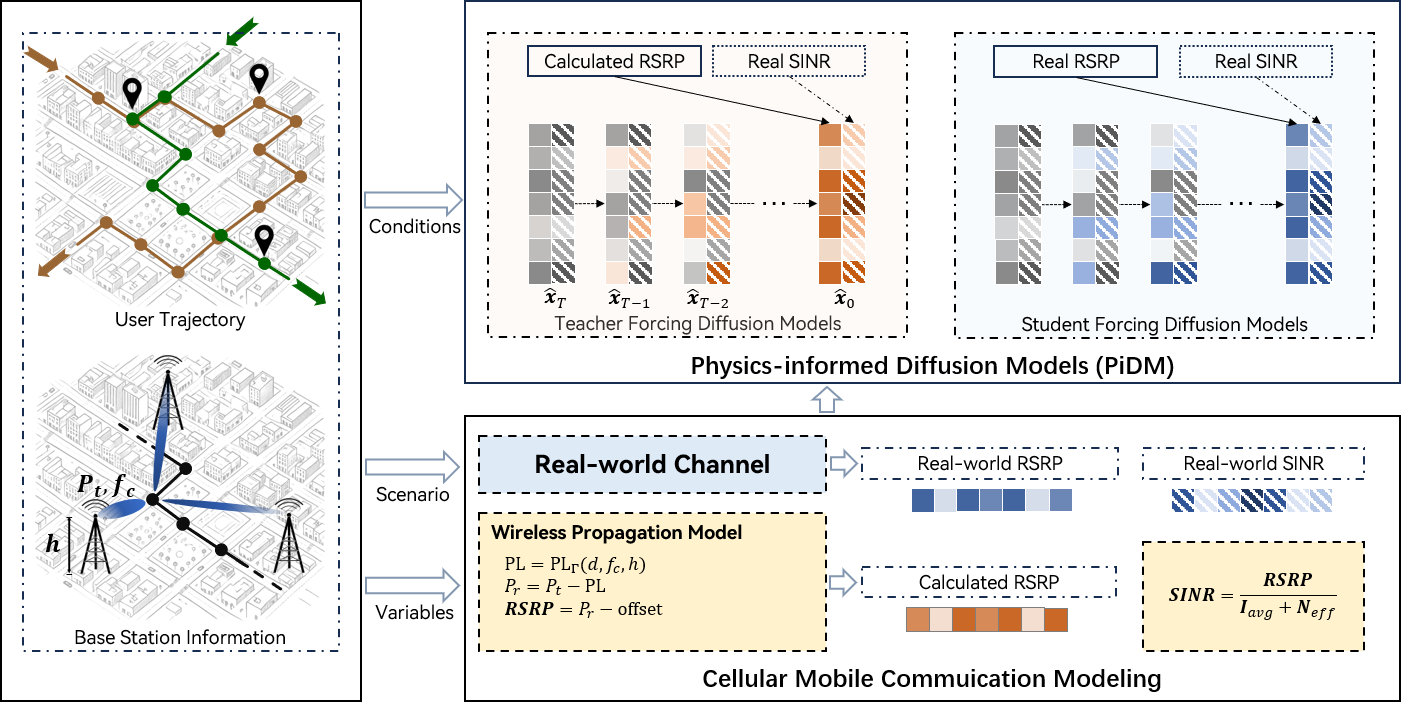}
\caption{Overall framework of the Physics-informed Diffusion Models for channel quality generation.}
\vspace{-4mm}
\label{Fig_syst}
\end{figure*}

\subsection{Teacher Model for Understanding physical Knowledge}
The Teacher Forcing Diffusion Model is shown on the left side of the top-right in Fig. \ref{Fig_syst}. The target data $\pmb{x}^{ta}$ is a two-dimensional vector of size $T \times 2$, where the first column $\pmb{x}^{ta}{[:,1]}$ represents RSRP, and the second column $\pmb{x}^{ta}{[:,2]}$ represents SINR. The training label $\hat{\pmb{x}}^{ta}$ is formed by concatenating the theoretical RSRP values $\hat{\textbf{RSRP}}_m^{1\times T}$ with the real SINR values $\textbf{SINR}_m^{1\times T}$.

For the Teacher Forcing phase, the loss function for model training is the \textbf{true correction}, defined as the Mean Squared Error (MSE) between the generated vector and the label $\hat{\pmb{x}}^{ta}$,
\begin{equation}
    \mathcal{L}_{\text{teacher}}=\text{MSE}(\hat{\pmb{x}}^{ta}_0,\pmb{x}^{ta}_t(\pmb{x}^{ta}_T,\epsilon))=||\hat{\pmb{x}}^{ta}_0-\pmb{x}^{ta}_t(\pmb{x}^{ta}_T,\epsilon)||^2_2.
\end{equation}

We used a 2D Transformer to extract data features. The Transformer is a neural network architecture originally designed for sequence-to-sequence tasks, primarily in natural language processing. Unlike traditional recurrent models, the Transformer employs a self-attention mechanism, which allows it to capture dependencies between all positions in the input sequence, regardless of their distance. The core component is the Scaled Dot-Product Attention, which computes attention weights as 
\begin{equation}
    \text{Attention}(Q, K, V) = \text{softmax}\left(\frac{QK^T}{\sqrt{d_k}}\right)V,
\end{equation}
where $ Q $ (query), $ K $ (key), and $ V $ (value) are projections of the input, and $ d_k $ is the dimension of the keys. This mechanism enables the model to focus on different parts of the input sequence when generating each element of the output, leading to better feature extraction.

The 2D Transformer is essentially a cascade of two Transformers, which encode and decode along the temporal dimension and the attribute dimension of $\pmb{x}^{ta}$, respectively. This design not only captures the temporal features of the sequence but also explores the relationship between RSRP and SINR. By doing so, it leverages the completeness of RSRP’s physical modeling to compensate for the limitations in SINR's physical modeling.

\subsection{Student Model for Capturing Environmental Features}
The Student Forcing Diffusion Model is shown on the left side of the top-right in Fig. \ref{Fig_syst}. The differences between Teacher Forcing train and Student Forcing training are in two aspects. First, the training label $\pmb{x}^{ta}$ is pure real data, formed by concatenating the real RSRP values $\hat{\textbf{RSRP}}_m^{1\times T}$ with the real SINR values $\textbf{SINR}_m^{1\times T}$. Second, the loss function for model training is a combination of \textbf{true correction} and \textbf{physical correction},
\begin{equation}
\begin{split}
    \mathcal{L}_{\text{teacher}}&=\gamma \cdot \text{MSE}(\pmb{x}^{ta}_0,\pmb{x}^{ta}_t(\pmb{x}^{ta}_T,\epsilon))+\delta \cdot \text{MSE}(\hat{\pmb{x}}^{ta}_0,\pmb{x}^{ta}_t(\pmb{x}^{ta}_T,\epsilon))\\
    &=\gamma \cdot ||\pmb{x}^{ta}_0-\pmb{x}^{ta}_t(\pmb{x}^{ta}_T,\epsilon)||^2_2 + \delta \cdot ||\hat{\pmb{x}}^{ta}_0-\pmb{x}^{ta}_t(\pmb{x}^{ta}_T,\epsilon)||^2_2,
\end{split}
\end{equation}
where $\gamma$ and $\delta$ are adjustable weights, and satisfy
\begin{equation}
    \gamma + \delta = 1.
\label{eq_tradeoff}
\end{equation}
\section{Experimental Results}
\subsection{Dataset}
Since user trajectories are typically considered private by operators, it is challenging for researchers to obtain complete datasets that simultaneously contain user trajectories, base station information, and channel quality. To experimentally verify the effectiveness of our model, we used the 5G Toolbox in MATLAB 2024b to construct several simulation scenarios based on real-world maps, thus generating datasets for machine learning. These maps represent a variety of environments, ranging from urban commercial districts to coastal rural areas. For each type of environment, we assigned different materials to reflectors such as buildings and ground surfaces to simulate the physical channels in different environments.
\begin{figure}[tb]
\centering
\includegraphics[width=0.8\linewidth]{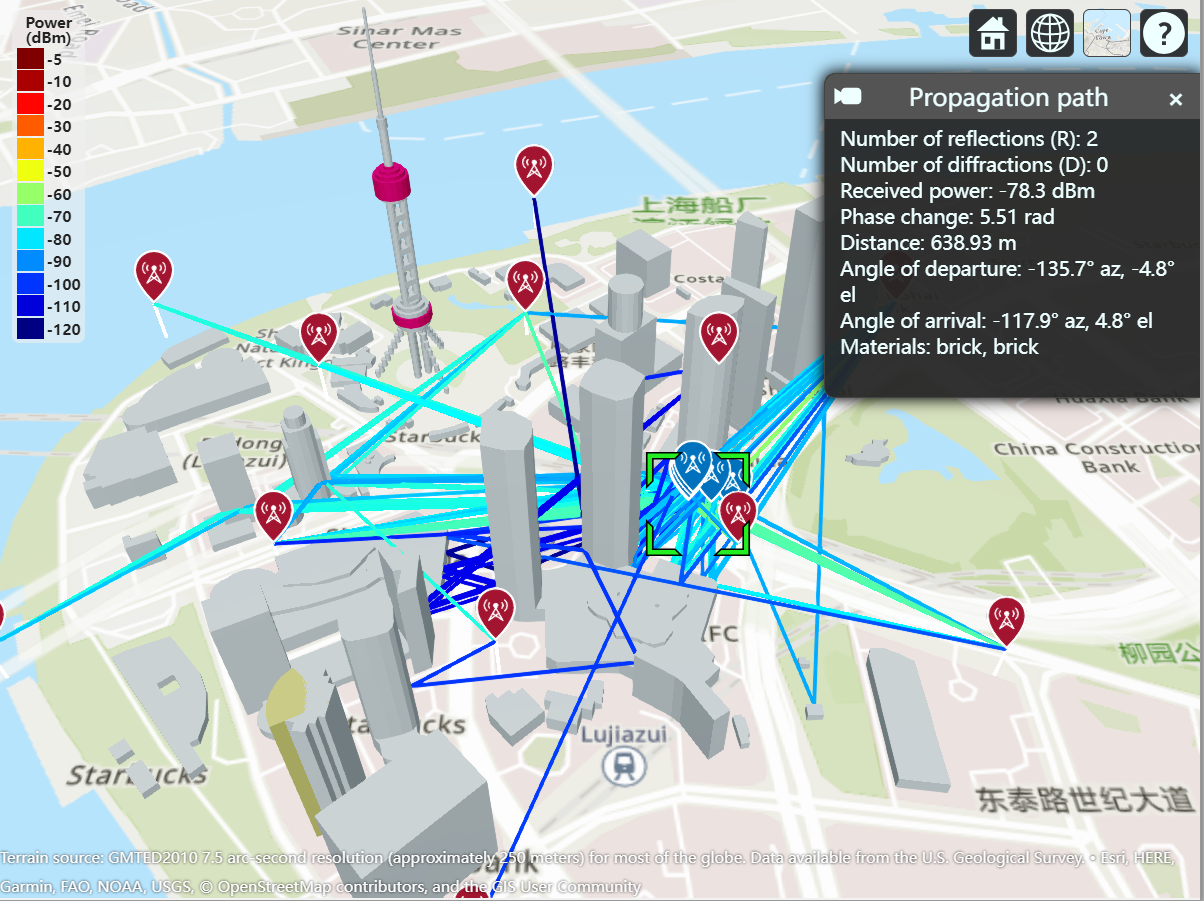}
\caption{Cellular mobile communication simulation in MATLAB.}
\vspace{-4mm}
\label{Fig_sim}
\end{figure}

The simulation scenario we constructed in the Lujiazui area of Pudong, Shanghai, is shown in Fig. \ref{Fig_sim}. In total, we generated 768 valid data samples from urban, suburban, and rural scenarios.


\subsection{Over Performance}
\begin{figure}[tb]
\centering
\includegraphics[width=\linewidth]{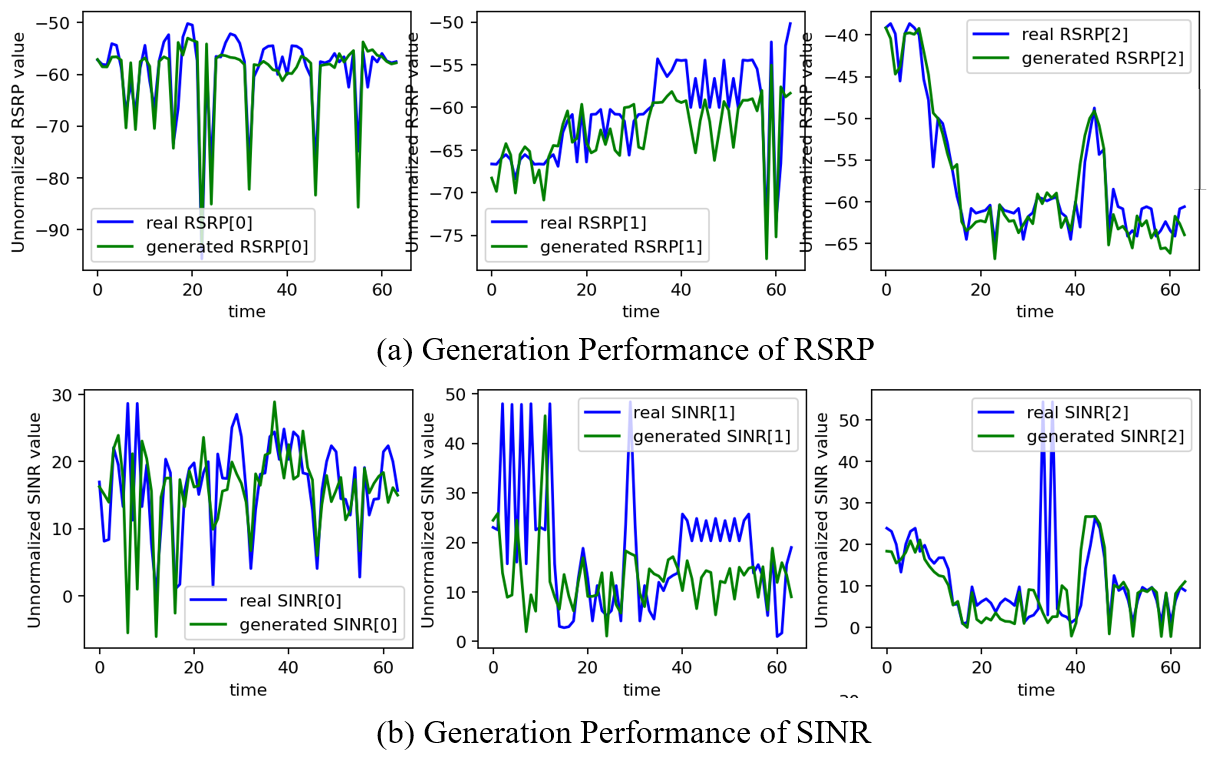}
\caption{Generation result of our model. The blue curve is the real data and the green curve is the generated data.}
\vspace{-4mm}
\label{Fig_main}
\end{figure}

Fig. \ref{Fig_main} shows the channel quality generation results on the simulation dataset using our proposed model. As can be seen, the model achieves high generation accuracy and fits well even in regions with significant fluctuations. We evaluated the results using three metrics: Jensen-Shannon Divergence (JSD), Total Variation (TV), and Normalized Root Mean Squared Error (NRMSE). The evaluation metrics is shown in Table \ref{Tab_main}.
\begin{table}[tb]
\centering
\small
\caption{RSRP and SINR generation performance. $\downarrow$ means lower is beter. Bold numbers denote the best results and \underline{underlined numbers} denote the second-best redults.}
\begin{tabular}{cccc}
\toprule
Metrics & JSD$\downarrow$ & TV$\downarrow$ & NRMSE$\downarrow$ \\
\midrule
RSRP & 0.0225 & 0.0623 & 0.1557 \\
SINR & 0.0663 & 0.0890 & 0.2093 \\
\bottomrule
\end{tabular}
\label{Tab_main}
\end{table}

The training iterations of the Teacher Forcing stage and the Student Forcing stage, as well as Eq. \ref{eq_tradeoff}, reflect a balance between physical knowledge and real data. We conducted a set of experiments to demonstrate the benefit of embedding physical knowledge into the diffusion model's generation performance, as well as to determine the optimal values of $\gamma$ and $\delta$. Table \ref{Tab_tradeoff} presents the results of this experiment. It can be observed that introducing physical knowledge through the Teacher Forcing stage improves the model's generation accuracy, and the balance between physical knowledge and real data achieves optimal performance when $\gamma=0.8$ and $\delta=0.2$.
\begin{table*}[tb]
\centering
\small
\caption{Experimental results for different sets of training stages and parameters. Metrics for RSRP, $\downarrow$ means lower is better.}
\begin{tabular}{lccccc}
\toprule
Stage × $N_{early\_stop}$ & $\gamma$ & $\delta$ & JSD$\downarrow$ & TV$\downarrow$ & NRMSE$\downarrow$ \\
\midrule
Student × 5 (T0S5) & 1 & 0 & 0.0265 & 0.0789 & 0.2029 \\
\midrule
\multirow{5}{*}{\shortstack{Teacher × 1 \\ Student × 4 (T1S4)}} & 1 & 0 & 0.0263 & 0.0789 & 0.2004 \\
& 0.95 & 0.05 & 0.0283 & 0.0824 & 0.2063 \\
& 0.9 & 0.1 & 0.0244 & 0.0792 & 0.1936 \\
& \textbf{0.8} & \textbf{0.2} & \textbf{0.0239} & \textbf{0.0737} & \textbf{0.1827} \\
& 0.6 & 0.4 & 0.0228 & 0.0754 & 0.1943 \\
\bottomrule
\end{tabular}
\label{Tab_tradeoff}
\end{table*}

\subsection{Few-shot and Zero-shot Learning}
\begin{figure*}[tb]
\centering
\includegraphics[width=\linewidth]{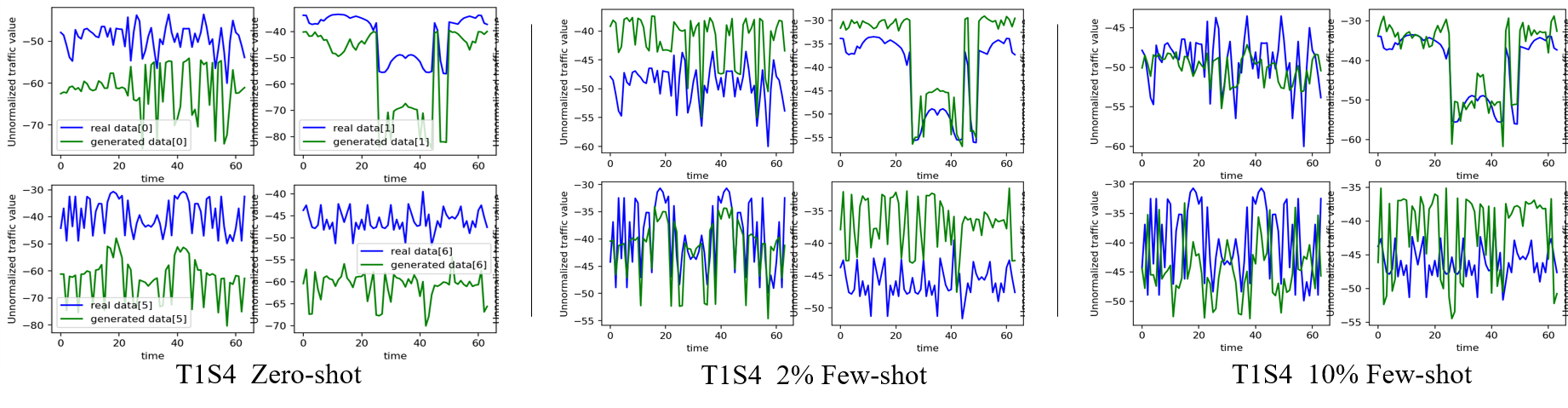}
\caption{Generated RSRP curves for few-shot and zero-shot learning. The blue curve is the real data and the green curve is the generated data.}
\vspace{-4mm}
\label{Fig_shot}
\end{figure*}

To verify the generalization ability of the model, we conducted few-shot and zero-shot experiments, with the results shown in Fig. \ref{Fig_shot} and Table \ref{Tab_shot}. It can be seen that the model achieves favorable generation performance under 10\% few-shot conditions.
\begin{table*}[tb]
\centering
\small
\caption{Experimental results for few-shot and zero-shot learning. Metrics for RSRP, $\downarrow$ means lower is better.}
\begin{tabular}{lcccccc}
\toprule
Stage × $N_{early\_stop}$ & Description & $\gamma$ & $\delta$ & JSD$\downarrow$ & TVD$\downarrow$ & NRMSE$\downarrow$ \\
\midrule
Student × 5 (T0S5) & Zero-shot & 1 & 0 & 0.0663 & 0.1098 & 0.7330 \\
& 2\% Few-shot & 1 & 0 & 0.0599 & 0.1152 & 0.3097 \\
& 10\% Few-shot & 1 & 0 & 0.0534 & 0.1096 & 0.2832 \\
\midrule
\multirow{3}{*}{\shortstack{Teacher × 1 \\ Student × 4 (T1S4)}} 
& Zero-shot & 0.8 & 0.2 & 0.0532 & 0.1026 & 0.6760 \\
& 2\% Few-shot & 0.8 & 0.2 & 0.0537 & 0.1221 & 0.2851 \\
& 10\% Few-shot & 0.8 & 0.2 & 0.0503 & 0.1083 & 0.2401 \\
\bottomrule
\end{tabular}
\label{Tab_shot}
\end{table*}

\section{Conclusion}
In conclusion, this paper presents a novel physics-informed diffusion model to address the limitations of existing methods for generating RSRP and SINR in cellular networks. By leveraging real environmental characteristics, base station configurations, and user trajectories, our model provides accurate predictions of wireless channel quality. The teacher-student framework used in this approach facilitates the integration of expert knowledge through knowledge distillation, significantly improving both the interpretability and accuracy of the model. Moreover, the modular and end-to-end design, combined with few-shot learning, enhances the model's adaptability across various scenarios, ensuring faster convergence and better generalization. These advancements offer valuable insights for optimizing base station deployment, predicting user network experience, and constructing real-world wireless communication simulators.



\bibliographystyle{IEEEtran}
\bibliography{Ref}

\vfill

\end{document}